\documentclass[5p,12pt]{elsarticle}
\usepackage[utf8]{inputenc}

\usepackage{amssymb}
\usepackage{amsmath}
\usepackage{mathtools}
\usepackage{bm}
\usepackage{enumerate}
\usepackage{listings}
\usepackage{color}
\usepackage[dvipsnames]{xcolor}
\usepackage{subcaption}
\usepackage{braket}
\PassOptionsToPackage{hyphens}{url}\usepackage{hyperref}
\biboptions{numbers,sort&compress} % This tells natbib to compress citation ranges, not done by default
\newcounter{bla}

\journal{a journal}
\definecolor{mygreen}{rgb}{0,0.6,0}
\definecolor{mygray}{rgb}{0.5,0.5,0.5}
\definecolor{mymauve}{rgb}{0.58,0,0.82}
\begin{document}

%===============================================================================
% Frontmatter
%===============================================================================
\begin{frontmatter}
\title{\texttt{DScribe}: Library of Descriptors for Machine Learning in Materials Science}

\author[aalto]{Lauri Himanen\corref{cor1}}
\ead{lauri.himanen@aalto.fi}
\author[aalto]{Marc O. J. J\"ager}
\author[aalto]{Eiaki V. Morooka}
\author[aalto,nanolayers]{Filippo Federici~Canova}
\author[aalto]{Yashasvi S. Ranawat}
\author[nanolayers,ntnu]{David Z. Gao}
\author[aalto,tu]{Patrick Rinke}
\author[aalto,mainz,kanazawa]{Adam S. Foster}

\cortext[cor1]{Corresponding author}
\address[aalto]{Department of Applied Physics, Aalto University, P.O. Box 11100, 00076 Aalto, Espoo, Finland}
\address[nanolayers]{Nanolayers Research Computing Ltd., 1 Granville Court, Granville Road, London, N12 0HL, United Kingdom}
\address[ntnu]{Department of Physics, Norwegian University of Science and Technology, NO-7491 Trondheim, Norway}
\address[mainz]{Graduate School Materials Science in Mainz, Staudinger Weg 9, 55128, Germany}
\address[kanazawa]{WPI Nano Life Science Institute (WPI-NanoLSI), Kanazawa University , Kakuma-machi, Kanazawa 920-1192, Japan}
\address[tu]{Theoretical Chemistry and Catalysis Research centre, Technische Universit\"at M\"unchen, Lichtenbergstr.~4, D-85747 Garching, Germany}

\begin{abstract}
DScribe is a software package for machine learning that provides popular feature transformations (\emph{``descriptors''}) for atomistic materials simulations. DScribe accelerates the application of machine learning for atomistic property prediction by providing user-friendly, off-the-shelf descriptor implementations. The package currently contains implementations for Coulomb matrix, Ewald sum matrix, sine matrix, Many-body Tensor Representation (MBTR), Atom-centered Symmetry Function (ACSF) and Smooth Overlap of Atomic Positions (SOAP). Usage of the package is illustrated for two different applications: formation energy prediction for solids and ionic charge prediction for atoms in organic molecules. The package is freely available under the open-source Apache License 2.0.
\end{abstract}


\end{frontmatter}

%===============================================================================
% Program summary
% \linenumbers

% Computer program descriptions should contain the following
% PROGRAM SUMMARY.
{\bf PROGRAM SUMMARY}

\begin{small}
\noindent
{\em Program Title: DScribe}                                           \\
{\em Licensing provisions: Apache-2.0}                                 \\
{\em Programming language: Python/C/C++}                               \\
{\em Supplementary material:} Supplementary Information as PDF         \\
{\em Nature of problem:} The application of machine learning for materials science is hindered by the lack of consistent software implementations for feature transformations. These feature transformations, also called descriptors, are a key step in building machine learning models for property prediction in materials science.\\
{\em Solution method:} We have developed a library for creating common descriptors used in machine learning applied to materials science. We provide an implementation the following descriptors: Coulomb matrix, Ewald sum matrix, sine matrix, Many-body Tensor Representation (MBTR), Atom-centered Symmetry Functions (ACSF) and Smooth Overlap of Atomic Positions (SOAP). The library has a python interface with computationally intensive routines written in C or C++. The source code, tutorials and documentation are provided online. A continuous integration mechanism is set up to automatically run a series of regression tests and check code coverage when the codebase is updated.
%{\em Additional comments including Restrictions and Unusual features (approx. 50-250 words):}  \\ %Provide any additional comments here.

%\begin{thebibliography}{0}
%\bibitem{1}Reference 1         % This list should only contain those items referenced in the
%\bibitem{2}Reference 2         % Program Summary section.
%\bibitem{3}Reference 3         % Type references in text as [1], [2], etc.
                                % This list is different from the bibliography at the end of
                                % the Long Write-Up.
%\end{thebibliography}
\end{small}

%===============================================================================
% Introduction
%===============================================================================
\section{Introduction}
\begin{figure*}[h!]
  \centering
  \includegraphics[width=0.95\textwidth]{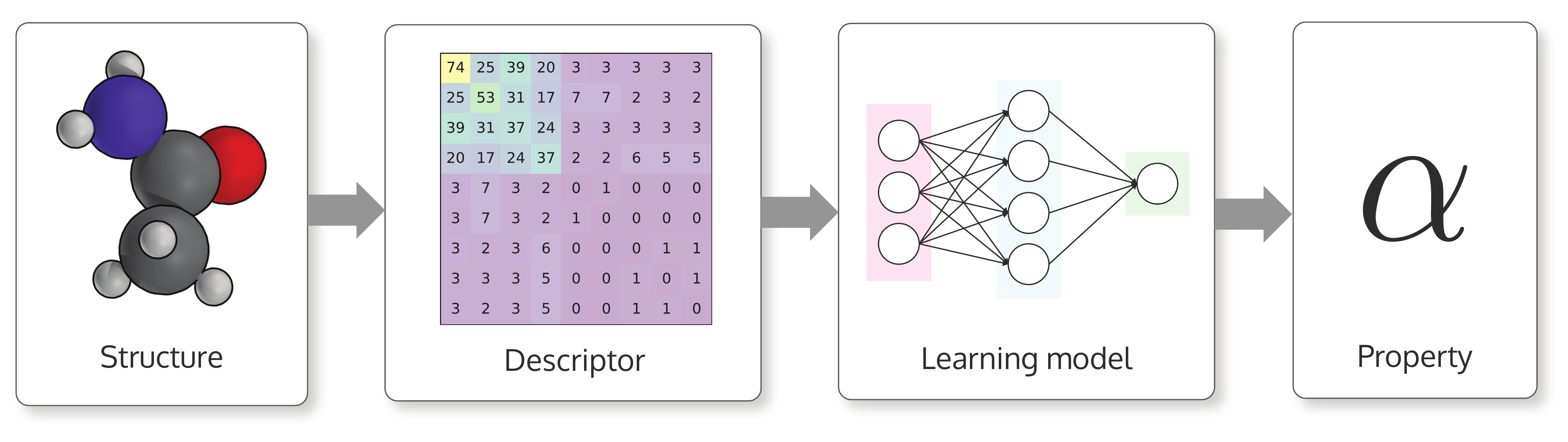}
  \caption{Typical workflow for making machine learning based materials property predictions for atomistic structures. An atomic structure is transformed into a numerical representation called a descriptor. This descriptor is then used as an input for a machine learning model that is trained to output a property for the structure. There is also a possibility of combining the descriptor and learning model together into one inseparable step.}
  \label{fig:workflow}
\end{figure*}
Machine learning of atomistic systems is a highly active, interdisciplinary area of research. The power of machine learning lies in the ability of interpolating existing calculations with surrogate models to accelerate predictions for new systems \cite{Takahashi:2016hr,Zdeborova:2017eo,Gubernatis:2018gi,Butler:2018fl}. The set of possible applications is very rich, including high-throughput search of stable compounds with machine learning based energy predictions for solids \cite{voronoi,mbtr,sm,Seko:2017dq}, accelerated molecular property prediction \cite{cm, Annika_KRR,Gosh/etal:2019}, creation of new force-fields based on quantum mechanical training data \cite{ani1,gap,soap,gdml,acsf,Behler:2016bp,Li:2017fm,Jacobsen:2018gh}, search for catalytically active sites in nanoclusters \cite{Li:2017ia,Goldsmith:2018bb,Chowdhury:2018ck,marcaki,Zahrt:2019jo} and efficient optimization of complex structures \cite{Kiyohara:2016bb,Zalake:2017gm,Milica_BOSS}.

%The traditional approach for calculating physical properties of an atomistic system is based on a computational model, such as density functional theory. If however there exists results from earlier computation, a machine learning model may be trained to provide a prediction for a new system.

%It is however important to keep in mind that ultimately machine-learning is interpolation and as such loses accuracy in the previously unexplored regions of the input space, possibly very quickly depending on the complexity of the underlying function. The success of this approach thus depends on the quantity that is predicted and the amount of training data that is available.

Atomistic machine learning establishes a relationship between the atomic structure of a system and its properties. This so called structure-property relation is illustrated in Fig.~\ref{fig:workflow}. It is analogous to the structure-property relation in quantum mechanics. For a set of nuclear charges $\{Z_i\}$ and atomic positions $\{R_i\}$ of a system, the solution of the Schr\"odinger equation $\hat{H}\Phi = E\Phi$ yields the properties of the system since both the Hamiltonian $\hat{H}$ and the wave function $\Phi$ depend only on $\{Z_i\}$ and $\{R_i\}$. Atomistic machine learning bypasses the computationally costly step of solving the Schr\"odinger equation\footnote{Of course in practice the Schr\"odinger equation is not solved directly but by approximate methods. Also approximate solutions of the Schr\"odinger equation are computationally expensive, compared to surrogate machine learning models.} by training a surrogate model. Once trained, the surrogate model is typically very fast to evaluate facilitating almost instant structure-property predictions.

Unlike for the Schr\"odinger equation, the nuclear charges and atomic positions are not a suitable input representation of atomistic systems for machine learning. They are, for example, not rotationally or translationally invariant. If presented with atomic positions, the machine learning method would have to learn rotational and translational invariance for every data set, which would significantly increase the amount of required training data. For this reason, the input data has to be transformed into a representation that is suitable for machine learning. This transformation step is often referred to as \emph{feature engineering} and the selected features are called a \textit{descriptor} \cite{lasso}\footnote{Sometimes the separation between a descriptor and a learning model is however blurred. For example, in various models based on neural networks \cite{cgcnn, dtnn, mpnn, schnet} the description of the atomistic structure is embedded inside the weights of the network, essentially mixing the descriptor and the learning model together.}. Various feature engineering approaches have been proposed \cite{cm, cm_versions, bob, sm, soap, mbtr, acsf, wacsf, voronoi, fragment, wyckoff, alchemy, f2bf3b, Seko:2017dq}, and often multiple approaches have to be tested to find a suitable representation for a specific task \cite{descreview1}. Features are often based on the atomic structure, but it is also common to extend the input to other system properties \cite{lasso, sisso, fragment, voronoi}.

There are several requirements for good descriptors in atomistic machine learning \cite{mbtr, sm}. We identify the following properties to be most important for an ideal descriptor:
\begin{enumerate}[i)]
  \item Invariant with respect to spatial translation of the coordinate system: isometry of space.
  \item Invariant with respect to rotation of the coordinate system: isotropy of space.
  \item Invariant with respect to permutation of atomic indices: changing the enumeration of atoms does not affect the physical properties of the system.
  \item Unique: there is a single way to construct a descriptor from an atomic structure and the descriptor itself corresponds to a single property.
  \item Continuous: small changes in the atomic structure should translate to small changes in the descriptor.
  \item Compact: the descriptor should contain sufficient information of the system for performing the prediction while keeping the feature count to minimum.
  \item Computationally cheap: the computation of the descriptor should be significantly faster than any existing computational model for directly calculating the physical property of interest.
\end{enumerate}

In this article we present the DScribe package that can be used to transform atomic structures into machine-learnable input features. The aim of this software is to provide a coherent and easily extendable implementation for atomistic machine learning and fast prototyping of descriptors. Currently in the DScribe package we include descriptors that can be represented in a vectorial form and are not dependent on any specific learning model. By decoupling the descriptor creation from the machine learning model, the user can experiment in parallel with various descriptor/model combinations and has the possibility of directly applying emerging learning models on existing data. This freedom to switch between machine learning models becomes important because currently no universally best machine model exists for every problem, as stated by the ``No Free Lunch Theorem'' \cite{nflt}. In practice this means that multiple models have to be tested to find optimal performance. Furthermore, vectorial features provide easier insight into the importance of certain features and facilitate the application of unsupervised learning methods, such as clustering and subsequent visualization with informative ``materials maps'' \cite{sketchmap, rematch, cartography}.

Descriptors that encode an atomic structure are typically designed to either depict a local atomic environment, or the structure as a whole. \emph{Global} descriptors encode information about the whole atomic structure. These global descriptors can be used to predict properties related to the structure as a whole, such as molecular energies \cite{cm}, formation energies \cite{voronoi} or band gaps \cite{fragment}. In this work we cover four such global descriptors: the Coulomb matrix \cite{cm}, the Ewald sum matrix \cite{sm}, the sine matrix \cite{sm} and the Many-Body Tensor Representation (MBTR) \cite{mbtr}. \emph{Local} descriptors are instead used to represent a localized region in an atomic structure, and are thus suitable for predicting localized properties, like atomic forces \cite{gap}, adsorption energies\cite{marcaki}, or properties that can be summed from local contributions. In this article we discuss two local descriptors, Atom-centered Symmetry functions (ACSFs) \cite{acsf} and the Smooth Overlap of Atomic Positions (SOAP) \cite{soap}.

We first introduce the descriptors that have been implemented in the DScribe package and then we discuss the structure and usage of the package. After this we illustrate the applicability of the package by showing results for formation energy prediction of periodic crystals and partial charge prediction for molecules. We conclude, by addressing the impact and future extensions of this package.

%===============================================================================
% Methods
%===============================================================================
\section{Descriptors}
Here we briefly introduce the different descriptors that are currently implemented in DScribe. In some cases, we have deviated from the original literature due to computational or other reasons, and if so this is explicitly mentioned. For more in-depth presentations of the descriptors we refer the reader to the original research papers. At the end of this section we also discuss methods for organizing the descriptor output so that it can be effectively used in typical machine learning applications.

\subsection{Coulomb matrix}
The Coulomb matrix \cite{cm} encodes the atomic species and inter-atomic distances of a finite system in a pair-wise, two-body matrix inspired by the form of the Coulomb potential. The elements of this matrix are given by:
\[
M^{\text{Coulomb}}_{ij} =
    \begin{cases}
      0.5Z_{i}^{2.4} & \forall~i = j\\
      \frac{Z_i Z_j}{\left\vert \bm{R}_i - \bm{R}_j\right\vert} & \forall~i \neq j\\
    \end{cases}
\]
where $Z$ is the atomic number, and $\left\vert \bm{R}_i - \bm{R}_j\right\vert$ is the Euclidean distance between atoms $i$ and $j$. The form of the diagonal terms was determined by fitting the potential energy of neutral atoms \cite{Ramakrishnan2015}.

\subsection{Ewald sum matrix}
\label{sec:ewald}
The Ewald sum matrix \cite{sm} can be viewed as a logical extension of the Coulomb matrix for periodic systems. In periodic systems each atom is infinitely repeated in the three crystal lattice vector directions, $\+a$, $\+b$ and $\+c$ and the electrostatic interaction between two atoms becomes
\begin{equation}
  \phi_{ij} = \sum_{\+n} \frac{Z_i Z_j}{\left\vert \bm{R}_i - \bm{R}_j\right\vert + \+n}
\end{equation}
where $\sum_{\+n}$ is the sum over all lattice vectors $\+n = h\+a + k\+b + l\+c$.

For $h, k, l \to \infty$, this sum converges only conditionally and will become infinite if the system is not charge neutral. In the Ewald sum matrix, the Ewald summation technique \cite{ewald1, ewald2} and a neutralizing background charge \cite{ewald3} is used to force this sum to converge. One can separate the total Ewald energy into pairwise components, which will result in the following matrix:
\[
M^{\text{Ewald}}_{ij} =
    \begin{cases}
      \phi^{\text{real}}_{ij} + \phi^{\text{recip}}_{ij} + \phi^{\text{self}}_{ij} + \phi^{\text{bg}}_{ij} & \forall~i = j \\
      2\left(\phi^{\text{real}}_{ij} + \phi^{\text{recip}}_{ij} + \phi^{\text{bg}}_{ij}\right) & \forall~i \neq j\\
    \end{cases}\\
\]
where the terms are given by
\begin{align}
  \label{eq:Ewald1}
  \phi_{ij}^{\text{real}} &= \frac{1}{2} Z_i Z_j \sum_{\bm{n'}} \frac{\mathrm{erfc}\left(\alpha \lvert \bm{R}_i - \bm{R}_j + \+n \rvert \right)}{\lvert \bm{R}_i - \bm{R}_j + \+n \rvert} \\
  \label{eq:Ewald2}
  \phi_{ij}^{\text{recip}} &= \frac{2\pi}{V} Z_i Z_j \sum_\+G \frac{e^{-\lvert \+G \rvert^2 / (2\alpha)^2}}{\lvert \+G \rvert^2} \cos \left( \+G \cdot \left(\bm{R}_i - \bm{R}_j\right)\right) \\
  \phi^{\text{self}}_{ij} &=
    \begin{cases}
      -\frac{\alpha}{\sqrt{\pi}} Z_i^2 & \forall~i = j\\
      0 & \forall~i \neq j\\
    \end{cases} \\
  \phi^{\text{bg}}_{ij} &=
    \begin{cases}
      -\frac{\pi}{2 V \alpha^2} Z_i^2 & \forall~i = j \\
      -\frac{\pi}{2 V \alpha^2} Z_i Z_j & \forall~i \neq j\\
    \end{cases}
\end{align}
Here the primed notation means that when $\bm{n} = \bm{0}$ the pairs $i = j$ are not taken into account. $\alpha$ is the screening parameter controlling the size of the gaussian charge distributions used in the Ewald method, $\+G$ is a reciprocal space lattice vector with an implicit $2\pi$ prefactor and $V$ is the volume of the cell. A more detailed derivation is given in the supplementary information. By default we use the value $\alpha = \sqrt{\pi}\left(\frac{N}{V^2}\right)^{1/6}$ \cite{ewald_opt}, where $N$ is the number of atoms in the unit cell.

It is important to notice that the off-diagonal contribution $\phi^{\text{self}}_{ij} + \phi^{\text{bg}}_{ij} = -\frac{\pi}{2 V \alpha^2} Z_i Z_j~\forall~i \neq j$ given here differs from the original work. In the original formulation this sum was defined as \cite{sm} $\phi^{\text{self}}_{ij} + \phi^{\text{bg}}_{ij} = -\frac{\alpha}{\sqrt{\pi}}(Z_i^2 + Z_j^2) -\frac{\pi}{2V\alpha^2}(Z_i + Z_j)^2~\forall~i \neq j$. Our correction makes the total matrix elements independent of the screening parameter $\alpha$, which is not the case in the original formulation.
\begin{figure*}[h]
  \centering
  \includegraphics[width=\textwidth]{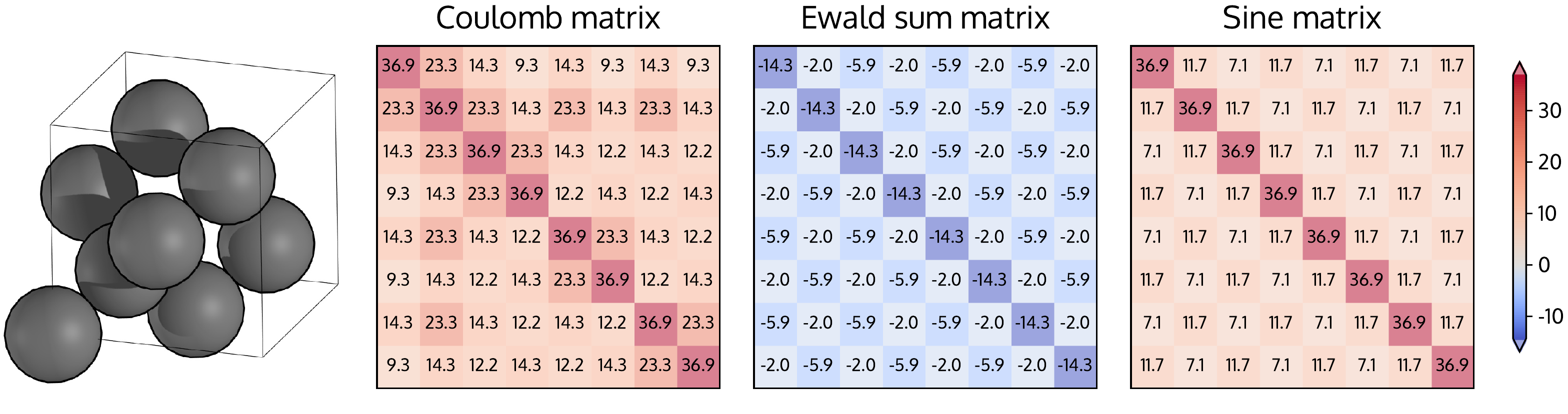}
  \caption{Illustration of the Coulomb matrix, Ewald sum matrix and sine matrix for a periodic diamond structure. The used atomic structure for the conventional diamond cell is shown on the left. The color scale (legend on the right) is used to illustrate the magnitude of the matrix elements.}
  \label{fig:matrices}
\end{figure*}

For numerical purposes, the sums in eqs.~\ref{eq:Ewald1} and \ref{eq:Ewald2} are cut off by  $n \leq n_\mathrm{cut}$ and $G \leq G_\mathrm{cut}$.  By default we use the values $G_\mathrm{cut} = 2\alpha\sqrt{-\ln A}$ and $n_\mathrm{cut} = \sqrt{-\ln A}/\alpha$\cite{ewald_opt}, where the small positive parameter $A$ controls the accuracy of the sum and can be determined by the user.

\subsection{Sine matrix}
The Ewald sum matrix encodes the correct Coulomb interaction between atoms, but can become computationally heavy especially for large systems. The sine matrix \cite{sm} captures some important features of interacting atoms in a periodic system with a much reduced computational cost. The matrix elements are defined by
\[
M^{\text{sine}}_{ij} =
    \begin{cases}
      0.5Z_{i}^{2.4} & \forall~i = j\\
      \phi_{ij} & \forall~i \neq j\\
    \end{cases}
\]
where
\begin{equation}
  \phi_{ij} = Z_i Z_j \lvert \+B \cdot \smashoperator{\sum\limits_{k=\{x, y, z\}}} \hat{\+e}_k \sin^2\left(\pi \+B^{-1} \cdot \left( \bm{R}_i - \bm{R}_j \right) \right) \rvert^{-1}
\end{equation}
Here $\+B$ is a matrix formed by the lattice vectors and $\hat{\+e}_k$ are the cartesian unit vectors. This functional form has no physical interpretation, but it captures some of the properties of the Coulomb interaction, such as the periodicity of the crystal lattice and an infinite energy when two atoms overlap.

The Coulomb, Ewald sum and sine matrices for diamond are depicted in Fig.~\ref{fig:matrices}. Notice that the matrices given here are not unique, as different cell sizes can be used for a periodic crystal, and the indexing of the rows and columns depends on the ordering of atomic indices in the structure. Section \ref{sec:local_to_global} discusses some ways to overcome the issues related to this non-uniqueness.

By construction the Coulomb matrix is not periodic as manifested by the unequivalent row elements in the matrix (one carbon in the system has four bonded neighbours, three carbons have two neighbours and four carbons have a single bonded neighbour). Conversely, both the Ewald sum and the sine matrix are periodic and correctly encode the identical environment of the carbon atoms in the lattice. As a result, each row and each column has the same matrix elements, but neighbouring rows and columns are shifted by one element relative to each other. Unlike the other matrices, Ewald sum matrix often contains negative elements due to the interaction of the positive atomic nuclei with the added uniform negative background charge. This energetically favourable interaction shifts the off-diagonal elements down in energy compared to the other two matrices. Moreover, the diagonal elements of the Ewald sum matrix encode the physical self-interaction of atoms with their periodic copies, instead of the potential energy of the neutral atoms.

\subsection{Many-body Tensor Representation}

The many-body tensor representation (MBTR) \cite{mbtr} encodes finite or periodic structures by breaking them down into distributions of differently sized structural motifs and grouping these distributions by the involved chemical elements. In MBTR, a geometry function $g_k$ is used to transform a configuration of $k$ atoms into a single scalar value representing that particular configuration. Our implementation provides terms up to $k=3$, and as the geometry functions uses $g_1(Z_l) = Z_l$ (atomic numbers), $g_2(\bm{R}_l, \bm{R}_m) = \frac{1}{\lvert \bm{R}_l - \bm{R}_m \rvert}$ (inverse distances) and $g_3(\bm{R}_l, \bm{R}_m, \bm{R}_n) = \cos (\angle (\bm{R}_l - \bm{R}_m, \bm{R}_n - \bm{R}_m))$ (cosines of angles). These scalar values are then broadened by using kernel density estimation with a gaussian kernel, leading to the following distributions $\mathcal{D}_k$
\begin{equation}
    \mathcal{D}_1^{l}(x) = \frac{1}{\sigma_1\sqrt{2 \pi}}e^{-\frac{\left(x-g_1\left(Z_l\right)\right)^2}{2\sigma_1^2}}
\end{equation}
\begin{equation}
    \mathcal{D}_2^{l,m}(x) = \frac{1}{\sigma_2\sqrt{2 \pi}}e^{-\frac{\left(x-g_2\left(\bm{R}_l, \bm{R}_m\right)\right)^2}{2\sigma_2^2}}
\end{equation}
\begin{equation}
    \mathcal{D}_3^{l,m,n}(x) = \frac{1}{\sigma_3\sqrt{2 \pi}}e^{-\frac{\left(x-g_3\left(\bm{R}_l, \bm{R}_m, \bm{R}_n\right)\right)^2}{2\sigma_3^2}}
\end{equation}
Here $\sigma_k$ is the standard deviation of the gaussian kernel and $x$ runs over a predefined range of values covering the possible values for $g_k$. A weighted sum of the distributions $\mathcal{D}_k$ are then made separately for each possible combination of $k$ chemical species present in the dataset. For $k=1,2,3$ these distributions are given by
\begin{align}
  \mathrm{MBTR}_1^{Z_1}(x) &= \sum_l^{\lvert Z_1\rvert} w_1^{l} \mathcal{D}_1^{l}(x) \label{eq:mbtr1} \\
  \mathrm{MBTR}_2^{Z_1,Z_2}(x) &= \sum_l^{\lvert Z_1\rvert}\sum_m^{\lvert Z_2\rvert} w_2^{l,m} \mathcal{D}_2^{l,m}(x) \label{eq:mbtr2} \\
  \mathrm{MBTR}_3^{Z_1,Z_2,Z_3}(x) &= \sum_l^{\lvert Z_1\rvert}\sum_m^{\lvert Z_2\rvert}\sum_n^{\lvert Z_3\rvert} w_3^{l,m,n}  \label{eq:mbtr3} \mathcal{D}_3^{l,m,n}(x)
\end{align}
where the sums for $l$, $m$ and $n$ run over all atoms with the atomic number $Z_1$, $Z_2$ or $Z_3$ respectively, and $w_k$ is a weighting function that is used to control the importance of different terms.  When calculating MBTR for periodic systems, the periodic copies of atoms in neighbouring cells are taken into account by extending the simulation cell periodically, and at least one of the indices $l, m, n$ has to belong to an atom in the original simulation cell.  Unlike in the original formulation \cite{mbtr}, we don't include the possible correlation between chemical elements directly in equations \eqref{eq:mbtr1}--\eqref{eq:mbtr3}. We don't however lose any generality, as the correlation between chemical elements can be introduced as a postprocessing step that combines information from the different species.

For $k=1$, typically no weighting is used: $w_1^l = 1$. In the case of $k=2$ and $k=3$, the weighting function can, however be used to give more importance to values that correspond to configuration where the atoms are closer together. For fully periodic systems, a weighting function must be used, as otherwise the sums in equations \eqref{eq:mbtr1}--\eqref{eq:mbtr3} do not converge. For $k=2,3$ we provide  exponential weighting functions of the form
\begin{align}
  w_2^{l,m} &= e^{-s_k \rvert \bm{R}_l - \bm{R}_{m}\rvert} \\
  w_3^{l,m,n} &= e^{-s_k \left( \rvert \bm{R}_l - \bm{R}_{m}\rvert + \rvert \bm{R}_m - \bm{R}_{n}\rvert  + \rvert \bm{R}_l - \bm{R}_{n}\rvert \right)}
\end{align}
where the parameter $s_k$ can be used to effectively tune the cutoff distance. For computational purposes a cutoff of $w_k^{\text{min}}$ can be defined to ignore any contributions for which $w_k < w_k^{\text{min}}$.

\begin{figure}[!h]
  \centering
  \includegraphics[width=0.48\textwidth]{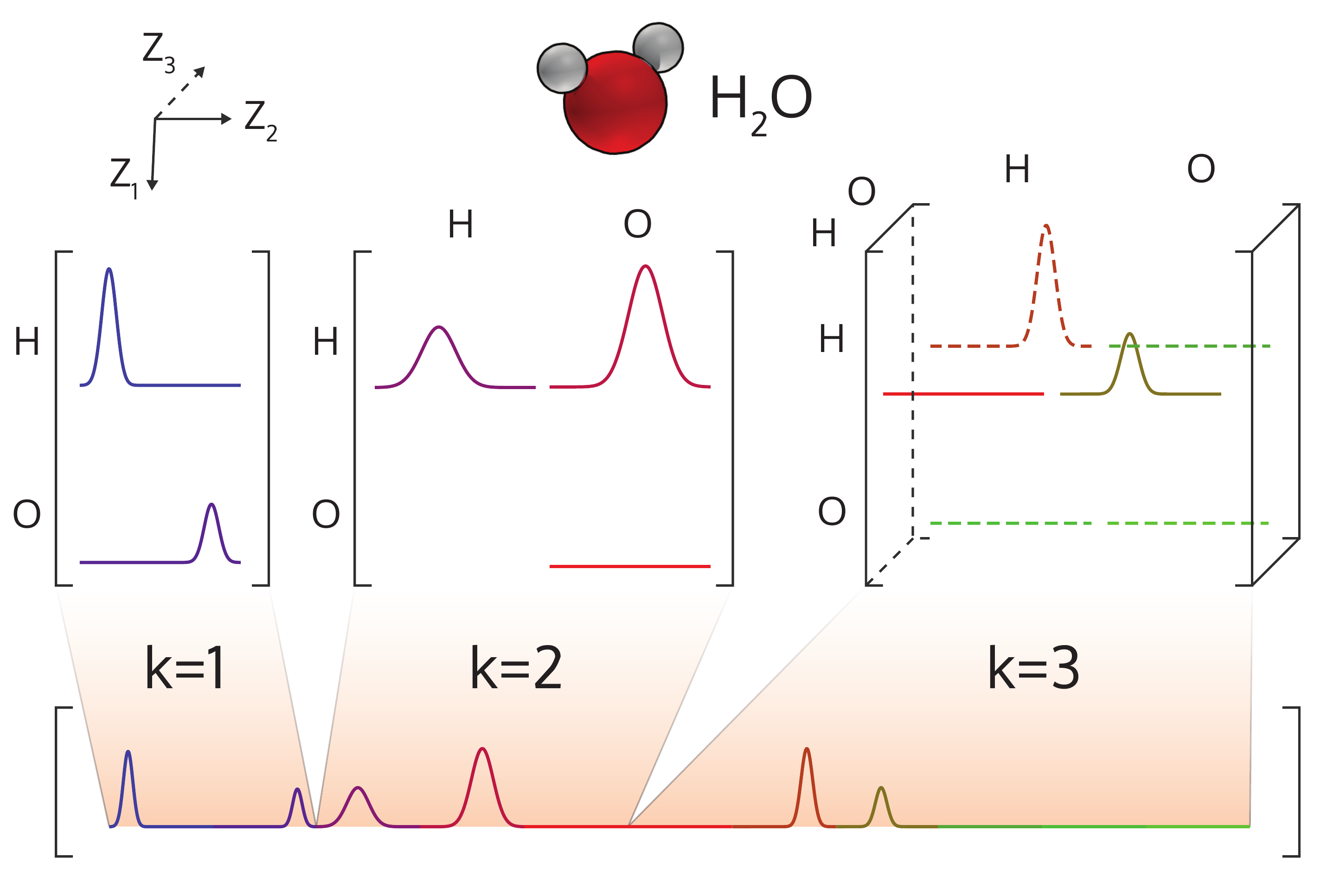}
  \caption{MBTR output for a water molecule showing the distributions $\mathrm{MBTR_k}$ for $k=1,2,3$ with different combinations of chemical elements. For each $k$-term, the distributions can be arranged into a $k$-dimensional grid, resulting in a $k+1$ dimensional tensor. If a flattened one-dimensional vector is needed by the learning model, the distributions may be concatenated together, possibly with some weighting, as shown in the lower panel.}
  \label{fig:mbtr}
\end{figure}

Some of the distributions, for example $\mathrm{MBTR}_2^{1, 2}$ and $\mathrm{MBTR}_2^{2, 1}$, contain identical information. In our implementation this symmetry is taken into consideration by only calculating the distributions for which the last atomic number is bigger or equal to the first: $Z_2 \geq Z_1$ in the case of $\mathrm{MBTR}_2^{Z_1, Z_2}$ or $Z_3 \geq Z_1$ in the case of $\mathrm{MBTR}_3^{Z_1, Z_2, Z_3}$. This reduces the computational time and the number of features in the final descriptor without losing information. The final MBTR output for a water molecule is illustrated in Fig. \ref{fig:mbtr}.

There are multiple system-dependent parameters that have to be decided for the MBTR descriptor. At each level $k$, the broadness of the distribution $\sigma_k$ has to be chosen. A too small value for $\sigma_k$ will lead to a delta-like distribution that is very sensitive to differences in system configurations. Conversely, a too large value will make it hard to resolve individual peaks as the distribution becomes broad and featureless. The choice of the weighting function also has a direct effect on the final distribution, as it controls how much importance is given to atom combinations that are physically far apart. When combining information from multiple $k$-terms, it can be beneficial to control the contribution from different terms. As the number of features related to higher $k$-values is bigger, the machine learning model may by default give more importance to these higher terms. For example if similarity between two MBTR outputs is measured by an Euclidean distance in the machine learning model, individually normalizing the total output for each term $k$ to unit length helps in equalizing the information content of the different terms.

\subsection{Atom-centered Symmetry Functions}
Atom-centered Symmetry Functions (ACSFs) \cite{acsf} can be used to represent the local environment near an atom by using a fingerprint composed of the output of multiple two- and three-body functions that can be customized to detect specific structural features. ACSF encodes the configuration of atoms around a single central atom with index $i$ by using so called symmetry functions. The presence of atoms neighbouring the central atom are detected by using three different two-body symmetry functions $G_i^{1,Z_1}$, $G_i^{2,Z_1}$ and $G_i^{3,Z_1}$, which are defined as

\begin{align*}
G_i^{1,Z_1} &= \sum_{j}^{\lvert Z_1 \rvert} f_\mathrm{c} \left( R_{ij} \right) \\
G_i^{2,Z_1} &= \sum_{j}^{\lvert Z_1 \rvert} e^{-\eta \left( R_{ij} - R_\mathrm{s} \right)^2} f_\mathrm{c}\left( r_{ij} \right) \\
G_i^{3,Z_1} &= \sum_{j}^{\lvert Z_1 \rvert} \cos\left( \kappa R_{ij} \right) f_\mathrm{c}\left( r_{ij} \right)
\end{align*}
where the summation for $j$ runs over all atoms with atomic number $Z_1$, $\eta$, $R_\mathrm{s}$ and $\kappa$ are user-defined parameters, $R_{ij} = \lvert\bm{R}_i-\bm{R}_j\rvert$ and $f_\mathrm{c}$ is a smooth cutoff function defined as
\begin{equation*}
f_\mathrm{c}\left( r \right) = \frac{1}{2} \left[ \cos\left( \pi \frac{r}{r_\mathrm{cut}} \right) + 1 \right]
\end{equation*}
where $r_\mathrm{cut}$ is a cutoff radius.

Additionally, three-body functions may be used to detect specific motifs defined by three atoms, one being the central atom. These three-body functions include a dependence on the angle between triplets of atoms within cutoff, as well as their mutual distance. The package implements the following functions

\begin{align}
\begin{split}
G_i^{4,Z_1,Z_2} &= 2^{1-\zeta} \sum_{j\neq i}^{\lvert Z_1 \rvert} \sum_{k\neq i}^{\lvert Z_2 \rvert} \left( 1+\lambda \cos\theta \right)^\zeta \\
 & \cdot e^{-\eta \left( R_{ij}^2 +R_{ik}^2+R_{jk}^2 \right)} f_\mathrm{c}\left( R_{ij} \right)f_\mathrm{c}\left( R_{ik} \right)f_\mathrm{c}\left( R_{jk} \right)
\end{split} \\
\begin{split}
G_i^{5,Z_1,Z_2} &= 2^{1-\zeta} \sum_{j\neq i}^{\lvert Z_1 \rvert} \sum_{k\neq i}^{\lvert Z_2 \rvert} \left( 1+\lambda \cos\theta \right)^\zeta \\
 & \cdot e^{-\eta \left( R_{ij}^2 +R_{ik}^2 \right)} f_\mathrm{c}\left( R_{ij} \right)f_\mathrm{c}\left( R_{ik} \right)
\end{split}
\end{align}
where the summation of $j$ and $k$ runs over all atoms with atomic numbers $Z_1$ or $Z_2$ respectively, $\zeta$, $\lambda$ and $\eta$ are user-defined parameters and $\theta$ is the angle between the three atoms ($i$-th atom in the center).
\begin{figure}
  \centering
  \includegraphics[width=0.49\textwidth]{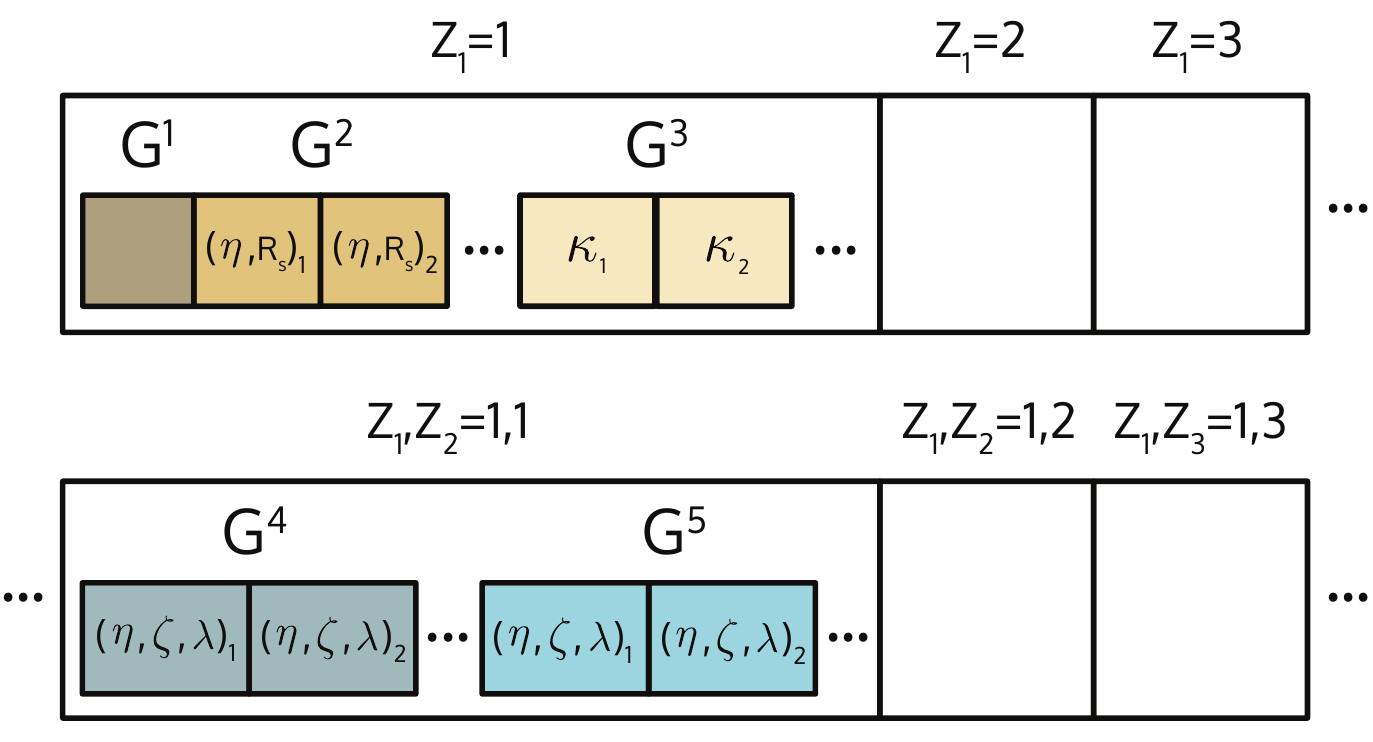}
  \caption{Structure of the ACSF output vector. The values of the two-body symmetry functions $G^1$, $G^2$ and $G^3$ are given first for each chemical species present in the dataset. Next the values of the three-body symmetry functions $G^4$ and $G^5$ are listed for each unique combination of two chemical species. All symmetry functions except $G^1$ may also have multiple parameterizations as indicated by the subindices.}
  \label{fig:acsfconcat}
\end{figure}

In practice, multiple symmetry functions of each type are used in the descriptor, each with a different parametrization ($\zeta$, $\lambda$, $\eta$, $R_\mathrm{s}$ and $\kappa$), encoding different portions of the chemical environment. While no optimal parameter set exists, the choice of parameters is guided by the generally desired properties of a descriptor previously listed (unique, continuous and compact), and various alternatives for different applications can be found in the literature \cite{acsf_appl_1, acsf_appl_2, acsf_appl_3}.

The final fingerprint for a single atom can be constructed by concatenating the output from differently parametrized symmetry functions with consistent ordering, as illustrated in Figure~\ref{fig:acsfconcat}. The list starts with the two-body ACSFs, ordered by atom type $Z_1$. For each type, $G^1$ appears first, bringing only one value since it has no parameter dependence. Next we find all values of $G^2$ calculated with different $(\eta,R_\mathrm{s})$ parameter pairs given by the user. The values of $G^3$ for all $\kappa$ are found last. This sequence is repeated for each atomic type, sorted from lighter to heavier. Three-body ACSFs appear afterward: for each unique combination of chemical elements, we find the values of $G^4$ and $G^5$ given by all specified triplets of $(\zeta,\lambda,\eta)$.

\subsection{Smooth Overlap of Atomic Orbitals}
The Smooth Overlap of Atomic Positions (SOAP) \cite{soap} can be used to encode a local environment within an atomic structure by using an expansion of a gaussian smeared atomic density based on spherical harmonics and radial basis functions. In SOAP, the atomic structure is first transformed into atomic density fields $\rho^Z$ for each species by using un-normalized gaussians centered on each atom
\begin{equation}
  \rho^{Z}(\bm{r}) = \sum_i^{\lvert Z \rvert}\,e^{-\frac{1}{2 \sigma^2} \lvert \bm{r} -\bm{R}_i \rvert ^2}.
\end{equation}
Here the summation for $i$ runs over atoms with the atomic number $Z$ to build a separate density for each atomic element and the width of the gaussian is controlled by $\sigma$.

A local region surrounding the point of interest set at the origin may then be expanded with a set of orthonormal radial basis functions and spherical harmonics as
\begin{equation}
  \rho^{Z}(\bm{r}) = \sum_{nlm} c^Z_{nlm} g_{n}(r)Y_{lm}(\theta, \phi)
\end{equation}
where the coefficients can be obtained through
\begin{equation}
  c^Z_{nlm}(\mathbf{x}) =\iiint_{\mathcal{R}^3}\mathrm{d}V g_{n}(r)Y_{lm}(\theta, \phi)\rho^Z(\bm{r}).
  \label{eq:soap_coeffs}
\end{equation}
For the spherical harmonics, an orthonormalized convention typical for quantum mechanics is used
\begin{equation}
  Y_{lm}(\theta, \phi) = (-1)^m\sqrt{\frac{(2l+1)}{4\pi}\frac{(l-m)!}{(l+m)!}}P_{lm}(\cos\theta)e^{im\phi}
\end{equation}
where $P_{lm}$ are the associated Legendre polynomials.

The final rotationally invariant output from our SOAP implementation is the partial power spectra \cite{rematch} vector $\mathbf{p}$ where the individual vector elements are defined as:
\begin{equation}
  p^{Z_1,Z_2}_{nn'l}
= \pi \sqrt{\frac{8}{2l+1}} \sum_m \left(c_{nlm}^{Z_1}\right)^*c^{Z_2}_{n'lm}
\end{equation}
The vector $\mathbf{p}$ is constructed by concatenating the elements $p^{Z_1,Z_2}_{nn'l}$ for all unique atomic number pairs $Z_1, Z_2$, all unique pairs of radial basis functions $n, n'$ up to $n_\mathrm{max}$ and the angular degree values $l$ up to $l_\mathrm{max}$.

Spherical harmonics are a natural orthogonal and complete set of functions for the angular degrees of freedom. For the radial degree of freedom the selection of the basis set is not as trivial and multiple approaches may be used. In our implementation we, by default, use a set of spherical primitive gaussian type orbitals $g_{nl}(r)$ as radial basis functions. These basis functions are defined as
\begin{align}
  g_{nl}(r) &= \sum_{n'=1}^{n_\mathrm{max}}\,\beta_{nn'l} \phi_{n'l}(r) \\
  \phi_{nl}(r) &= r^l e^{-\alpha_{nl}r^2}.
  \label{eq:gto}
\end{align}
This basis set allows analytical integration of the $c_{nlm}$ coefficients defined by equation \eqref{eq:soap_coeffs}. This provides a speedup over various other radial basis functions that require numerical integration. Our current implementation provides the analytical solutions up to $l \leq 9$, with the possibility of adding more in the future.

\begin{figure}[h!]
  \centering
  \includegraphics[width=0.48\textwidth]{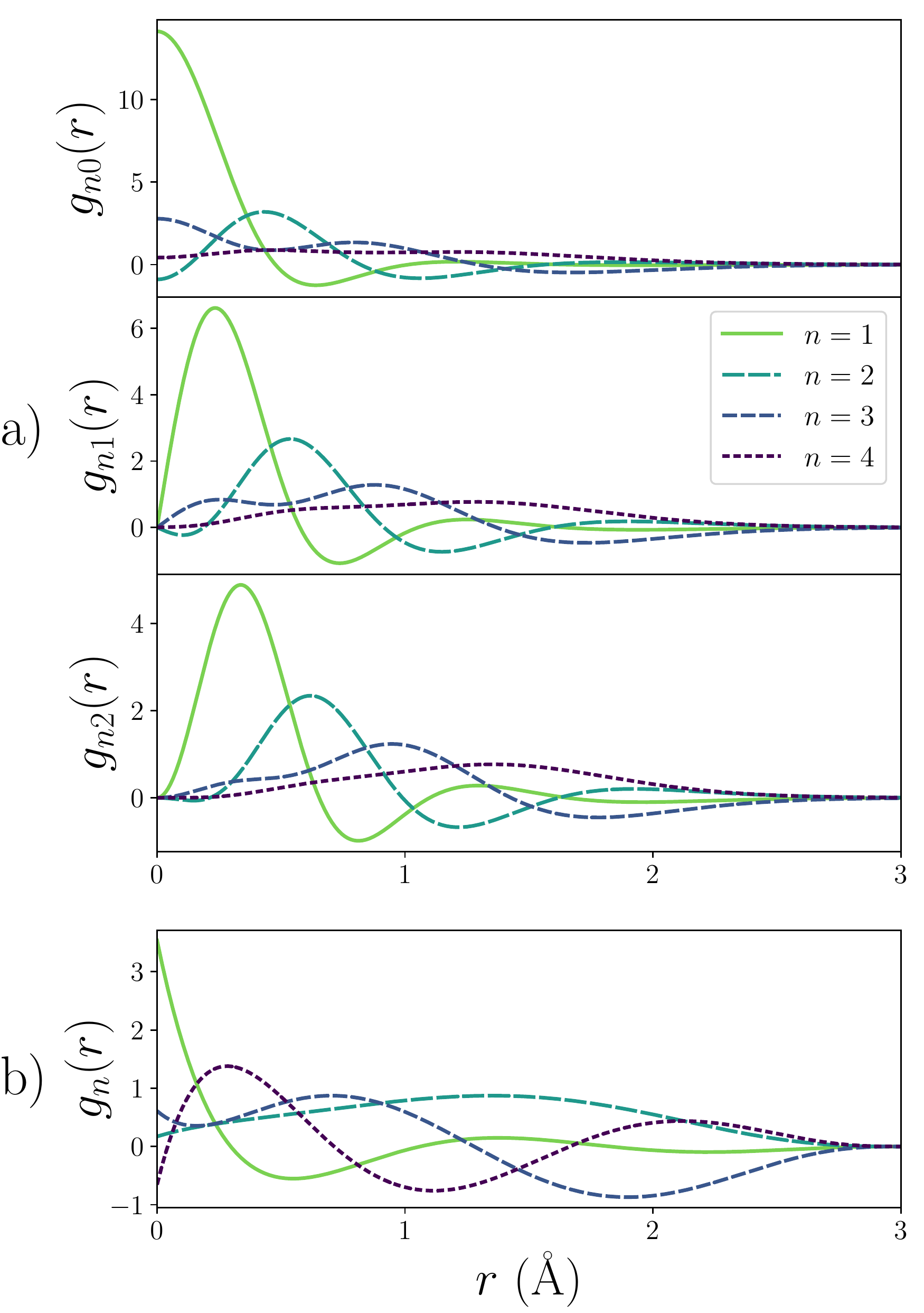}
  \caption{Plot of the a) gaussian type orbital and b) polynomial radial basis functions, defined by equations \eqref{eq:gto} and \eqref{eq:poly} respectively. The basis functions here correspond to the orthonormalized set with $r_\mathrm{cut} = 3$ and up to $n_\mathrm{max}=4$. Notice that the polynomial basis is independent of the spherical harmonics degree $l$, whereas the form of the gaussian type orbital basis depends on $l$ and the examples here are given for $l=0,1,2$.}
  \label{fig:soapbasis}
\end{figure}

The decay parameters $\alpha_n$ are chosen so that each non-orthonormalized function $\phi_{nl}$ decays to a threshold value of $10^{-3}$ at a cutoff radius taken on an evenly spaced grid from 1Å to $r_\mathrm{cut}$ with a step of $\frac{r_\mathrm{cut}-1}{n_\mathrm{max}}$. Thus the parameter $r_\mathrm{cut}$ controls the maximum reach of the basis and a better sampling can be obtained by increasing the number of basis functions $n_\mathrm{max}$.

The weights $\beta_{nn'l}$ are chosen so that the radial basis functions are orthonormal. For each value of angular degree $l$, the orthonormalizing weights $\beta_{nn'l}$ can be calculated with Löwdin orthogonalization \cite{lowdin}:
\begin{align}
  \bm{\beta} &= \bm{S}^{-1/2} \\
  S_{nn'} &= \braket{\phi_{nl}|\phi_{n'l}} = \int_0^\infty \mathrm{d}r r^2 r^l e^{-\alpha_{nl}r^2} r^l e^{-\alpha_{n'l}r^2}
  \label{eq:normalization}
\end{align}
where $\bm{S}$ is the overlap matrix and $\bm{\beta}$ is a matrix containing the weights $\beta_{nn'l}$.

We also provide an option for using the radial basis consisting of cubic and higher order polynomials, as introduced in the original SOAP article \cite{soap}. This basis set is defined as:
\begin{align}
  g_{n}(r) &= \sum_{n'=1}^{n_\mathrm{max}}\,\beta_{nn'} \phi_{n'}(r) \\
  \phi_{n}(r) &= (r-r_\mathrm{cut})^{n+2}
  \label{eq:poly}
\end{align}
The calculations with this basis are performed with efficient numerical integration and currently support $l_\mathrm{max} \leq 20$.

The two different basis sets are compared in Figure \ref{fig:soapbasis}. Most notably the form of the gaussian type orbitals depend on the angular degree $l$, whereas the polynomial basis is independent of this value. It is also good to notice that between these two radial basis functions the definition of $r_\mathrm{cut}$ is somewhat different -- whereas the polynomial basis is guaranteed to decay to zero at $r_\mathrm{cut}$, the gaussian basis only approximately decays near this value and the decay is also affected by the orthonormalization.

%The limitation compared to the polynomial basis, as introduced in the original article, is that this basis does not offer such an even coverage of the radial features, but will more accurately reconstruct features that are closer to the atoms.

\subsection{Descriptor usage as machine learning input}
\label{sec:local_to_global}

In this section we discuss some of the methods for organizing the output from descriptors so that it can be efficiently used as input for machine learning.

The descriptor invariance against permutations of atomic indices -- property iii) in the introduction -- is directly achieved in MBTR, SOAP and ACSF by stratifying the output according to the involved chemical elements. The output is always ordered by a predefined order determined by the chemical elements that are included in the dataset, making the output independent of the indexing of individual atoms. The three matrix descriptors -- the Coulomb matrix, Ewald sum matrix, and sine matrix -- are, however, not invariant with respect to permutation of atomic indices, as the matrix columns and rows are ordered by atomic indices. However, there are different approaches for enforcing invariance for these matrices. One way is to encode the matrices by their eigenvalues, which are invariant to changes in the column and row ordering \cite{cm}. Another way is to order the rows and columns by a chosen norm, typically the Euclidean norm \cite{cm_versions}. A third approach is to augment the dataset by creating multiple slightly varying matrices for each structure. In this approach multiple matrices are drawn from a statistical set of sorted matrices where Gaussian noise is added to the row norms before sorting \cite{cm_versions}. When the learning algorithm is trained over this ensemble of matrices it becomes more robust against small sorting differences that can be considered noise. All of these three approaches are available in our implementation.

Machine learning algorithms also often require constant-sized input. Once again the stratification of the descriptor output by chemical elements makes the output for MBTR, ACSF and SOAP constant size. For the matrix descriptors a common way to achieve a uniform size for geometries with different amount of atoms, is by introducing zero-padding. This means that we first have to determine the largest system in the dataset. If this system has $N_\mathrm{max}$, we allocate matrices of size $N_\mathrm{max} \times N_\mathrm{max}$ or a vectors or size $N_\mathrm{max}$ if using matrix eigenvectors. The descriptor for each system will fill the first $N^2$ or $N$ many entries, with the rest being set to zero. If the machine-learning algorithms expects a one-dimensional vector as input, the two-dimensional matrices can be flattened by concatenating the rows together into a single vector.

Local descriptors, such as ACSF and SOAP, encode only local spatial regions and cannot be directly used as input for performing predictions related to entire structures. There are, however, various ways for combining information from multiple local sites to form a prediction for an entire structure. The descriptor output for multiple local sites can simply be averaged, a custom kernel can be used to combine information from multiple sites \cite{soap_alchemy_multi, rematch} or the predicted property can in some cases be directly modeled as a sum of local contributions \cite{gap}.

%===============================================================================
% Software structure
%===============================================================================
\section{Software structure}
We use python as the default interfacing language through which the user interacts with the library. This decision was motivated by the existence of various python libraries, including ase \cite{ase-paper}, pymatgen \cite{pymatgen} and quippy \cite{quippy}, that supply tools for creating, reading, writing and manipulating atomic structures. Our python interface does not, however, restrict the implementation to be made entirely in python. Python can easily interact with software libraries written with high-performance, statically typed languages such as C, C++ and Fortran. We use this dual approach by performing some of the most computationally heavy calculations either in C or C++.

\begin{figure}[h]
    \centering
\begin{lstlisting}[language=Python]
from ase.build import molecule
from dscribe.descriptors import SOAP

# Define the atomic system
system = molecule("H2O")

# Setup the descriptor
soap = SOAP(
    species=["H", "O"],
    rbf="gto",
    rcut=3,
    nmax=10,
    lmax=8,
    sparse=True
)

# Query the number of features
n_features = soap.get_number_of_features()

# Get output as a dense or sparse array
output = soap.create(system)

# Get output for multiple systems
systems = [molecule("H2"), molecule("O2")]
outputs = soap.create(systems, n_jobs=2)
\end{lstlisting}
    \caption{Example of creating descriptors with DScribe. The structures are defined as ase.Atoms objects, in this case by using predefined molecule geometries. The usage of all descriptors follows the same pattern: a) a descriptor object is initialized with the desired configuration b) the number of features can be requested with \texttt{get\_\-number\_of\-\_features} c) the actual output is created with \texttt{create}-method that takes one or multiple atomic structures and possibly other arguments, such as the number of parallel jobs to use.}
    \label{fig:interface}
\end{figure}

An example of creating a descriptor for an atomic structure with the library is demonstrated in Fig. \ref{fig:interface}. It demonstrates the workflow that is common to all descriptors in the package. For each descriptor we define a class, from which objects can be instantiated with different descriptor specific setups.

All the descriptors have the \texttt{sparse}-parameter that controls whether the created output is a dense or a sparse matrix. The possibility for creating a sparse output is given so that large and sparsely filled output spaces can be handled, as typically encountered when a dataset contains large amounts of different chemical elements. Various machine learning algorithms can make use of this sparse matrix output with linear algebra routines specifically designed for sparse data structures.

Once created, the descriptor object is ready to be used and provides different methods for interacting with it. All of the descriptors implement two methods: \texttt{get\_\-number\_of\-\_features} and \texttt{create}. The \texttt{get\_\-number\_of\-\_features}-method can be used for querying the final number of features for the descriptor, even before a structure has been provided. This dimension can be used for initializing and reserving storage space for the resulting output array. \texttt{create} accepts one or multiple atomistic structures as an argument, and possibly other descriptor-specific arguments. It returns the final descriptor output that can be used in machine learning applications. To define atomic structures we use the ase.Atoms-object from the ase package\cite{ase-paper}. The Atoms-objects are easy to create from structure files or build with the utilities provided by ase.

As the creation of a descriptor for an atomic system is completely independent from the other systems, it can be easily parallelized with data parallelism. For convenience we provide a possibility of parallelizing the descriptor creation for multiple samples over multiple processes. This can be done by simply providing the number of parallel jobs to instantiate with the \texttt{n\_jobs}-para\-me\-ter as demonstrated in Figure \ref{fig:interface}.

The DScribe package is structured such that new descriptors can easily be added. We provide a python base-class that defines a standard interface for the descriptors through abstract classes. One of our design goals is to provide a codebase in which researchers can make their own descriptors available to the whole community. All descriptor implementations are accompanied by a test module that defines a set of standard tests. These tests include tests for rotational, translational and index permutation invariance, as well as other tests for checking the interface and functionality of the descriptor. We have adapted a continuous integration system that automatically runs a series of regression tests when changes in the code are introduced. The code coverage is simultaneously measured as a percentage of visited code lines in the python interface.

The source code is directly available in github at \url{https://github.com/SINGROUP/dscribe} and we have created a dedicated home page at \url{https://singroup.github.io/dscribe/} that provides additional tutorials and a full code documentation. For easy installation the code is provided through the python package index (pip) under the name \texttt{dscribe}.

%===============================================================================
% Results and discussion
%===============================================================================
\section{Results and discussion}
The performance of the DScribe package is illustrated for formation energy predictions for inorganic crystal structures and ionic charge prediction for atoms in organic molecules. These examples demonstrate the usage of the package in supervised machine learning tasks, but the output vectors can be as easily used in other learning tasks. For example the descriptors can be used as input for unsupervised clustering algorithms such as T-distributed stochastic neighbor embedding (T-SNE) \cite{tsne} or Sketchmap \cite{sketchmap} to analyse structure-property relations in structural and chemical landscapes.

For simplicity we here restrict the machine learning model to be kernel ridge regression (KRR) as implemented in the scikit-learn \cite{scikit-learn}-package. However, the vectorial nature of the output from all the introduced descriptors does not impose any specific learning scheme, and many other regressors can be used, including neural networks, decision trees and support vector regression.

\subsection{Formation energy prediction for inorganic crystals}
\begin{figure}[h]
  \centering
  \includegraphics[width=0.48\textwidth]{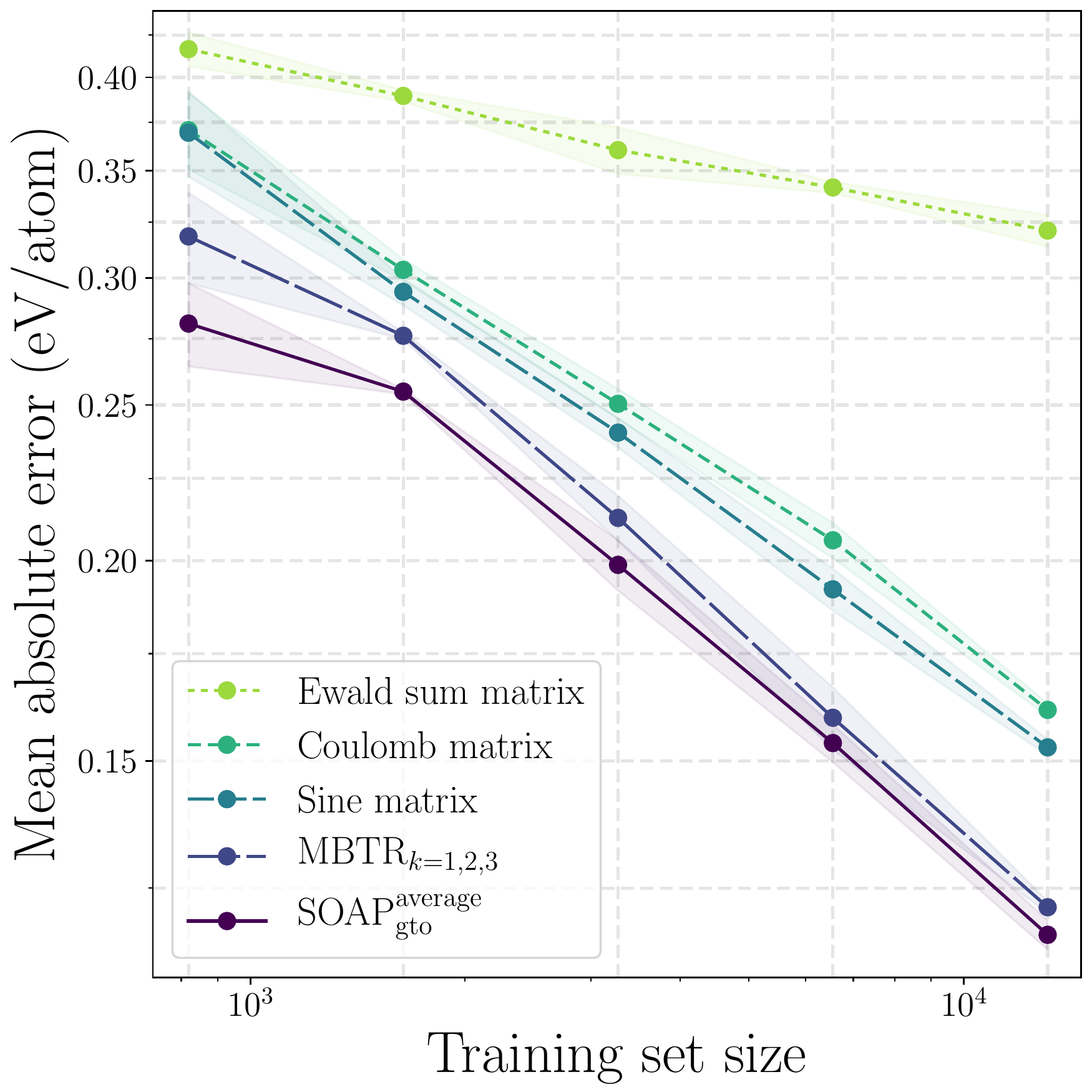}
  \caption{Mean absolute error for formation energy predictions as a function of training set size. The data consists of inorganic crystals from the OQMD database. The predictions are performed with KRR and five different descriptors: Ewald sum matrix, Coulomb matrix, sine matrix, MBTR$_{k=1,2,3}$ and an averaged SOAP output for all atoms in the crystal. The figure shows an average over three randomly selected datasets, with the standard deviation shown by the shaded region.}
  \label{fig:formation_energies}
\end{figure}
\begin{figure}[h]
  \centering
  \includegraphics[width=0.48\textwidth]{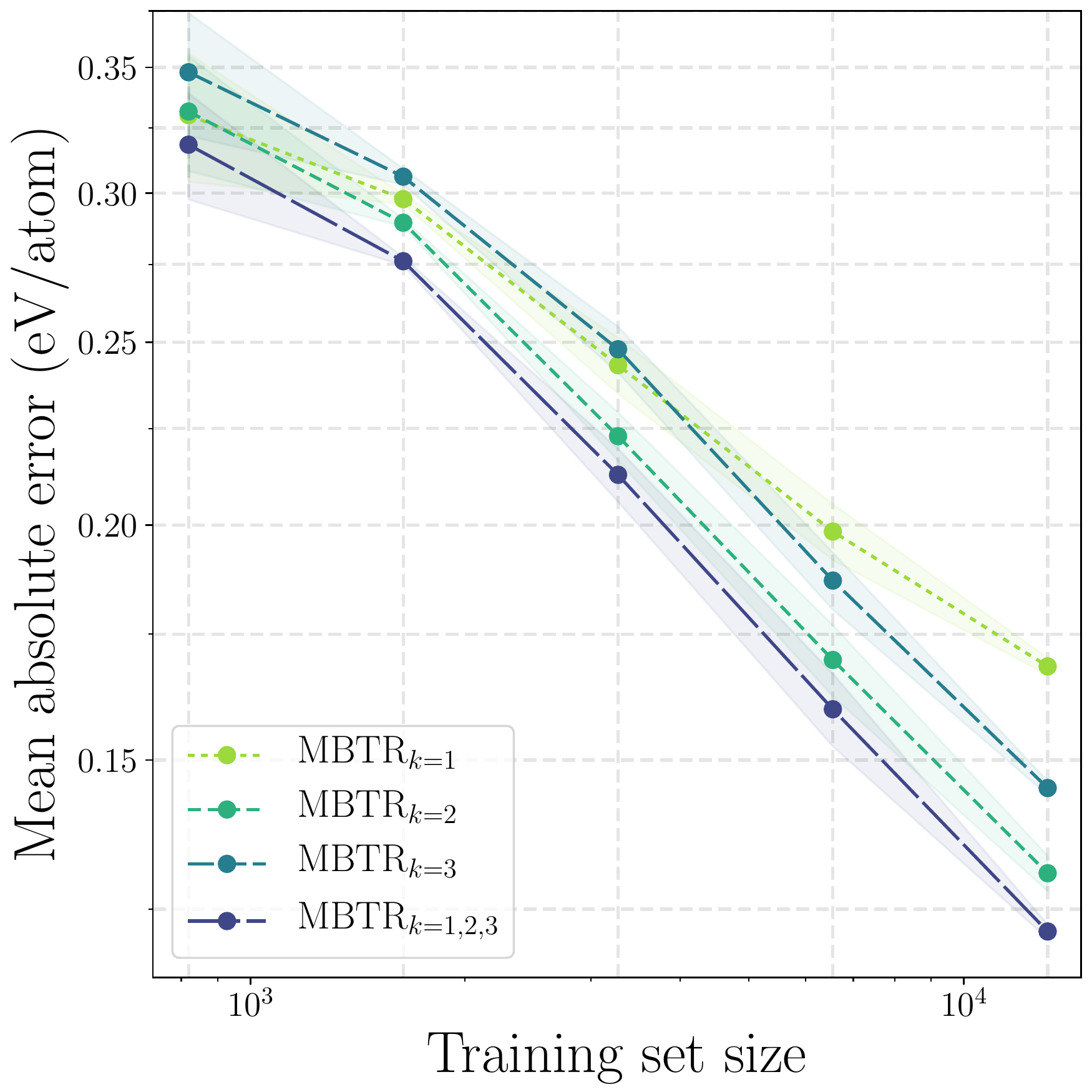}
  \caption{Breakdown of the formation energy prediction error for different MBTR-terms. The predictions are performed with KRR and four different MBTR configurations: MBTR$_{k=1}$, MBTR$_{k=2}$, MBTR$_{k=3}$ and MBTR$_{k=1,2,3}$ which includes all three terms, each term normalized to unit length.  The figure shows an average over three randomly selected datasets, with the standard deviation shown by the shaded region.}
  \label{fig:formation_energies_mbtr}
\end{figure}

We demonstrate the use of multiple descriptors on the task of predicting the formation energy of inorganic crystals. The data comes from the Open Quantum Materials Database (OQMD) 1.1 \cite{oqmd}. We selected structures with a maximum of 10 atoms per unit cell and a maximum of 6 different atomic elements. Unconverged systems were filtered by removing samples which have a formation energy that is more than two standard deviations away from the mean, resulting in the removal of 96 samples. After these selections, 222 215 samples were left. The predictions were performed on randomly selected subsets with sizes 1024, 2048, 4096, 8192 and 16384. The mean absolute errors as a function of training set size are given in Figure \ref{fig:formation_energies}.

The Coulomb matrix, Ewald sum matrix and sine matrix are used for the prediction with matrix rows and columns sorted by their Euclidean norm, and using the unit cell that was used for performing the formation energy calculation. The Coulomb matrix does not take the periodicity of the structure into account, but is included as a baseline for the other methods. We include MBTR with different values of $k$ and for each $k$ we individually optimize $\sigma$ and $s_k$ with grid search. Figure \ref{fig:formation_energies_mbtr} shows the error for each tested MBTR term, and the best performing one is included in Figure \ref{fig:formation_energies}. To test the energy prediction by combining information from multiple local descriptors, as discussed in \ref{sec:local_to_global}, we also include results using a simple averaged SOAP output for all atoms in the simulation cell. For SOAP we use the gaussian type orbital basis and fix $n_\mathrm{max} = 8$ and $l_\mathrm{max} = 8$, but optimize the cutoff $r_\mathrm{cut}$ and gaussian width $\sigma$ individually with grid search.

The possible descriptor hyperparameters are optimized at a subset of $2^{12} = 4096$ samples with 5-fold cross-validation and 80\%/20\%-training/test split. The KRR kernel width and the regularization parameter are also allowed to vary on a logarithmic grid during the descriptor hyperparameter search. The use of a smaller subset allows much quicker evaluation for the hyperparameters than optimizing the hyperparameters for each size individually, but the transferability of these optimized hyperparameters to different sizes may affect the results slightly.

After finding the optimal descriptor setup, it is used to train a model on dataset sizes of 1024, 2048, 4096, 8192 and 16384. The same cross-validation setup as for the descriptor hyperparameter optimization is used, but now with finer grid for the KRR kernel width. The predictions are repeated three times on different training sets to estimate the variance of the results. The hyperparameter grids and optimal values for both the descriptors and kernel ridge regression are found in the Supplementary Information together with additional details.

\subsection{Ionic charge prediction for organic molecules}

\begin{figure*}[h]
  \centering
  \includegraphics[width=\textwidth]{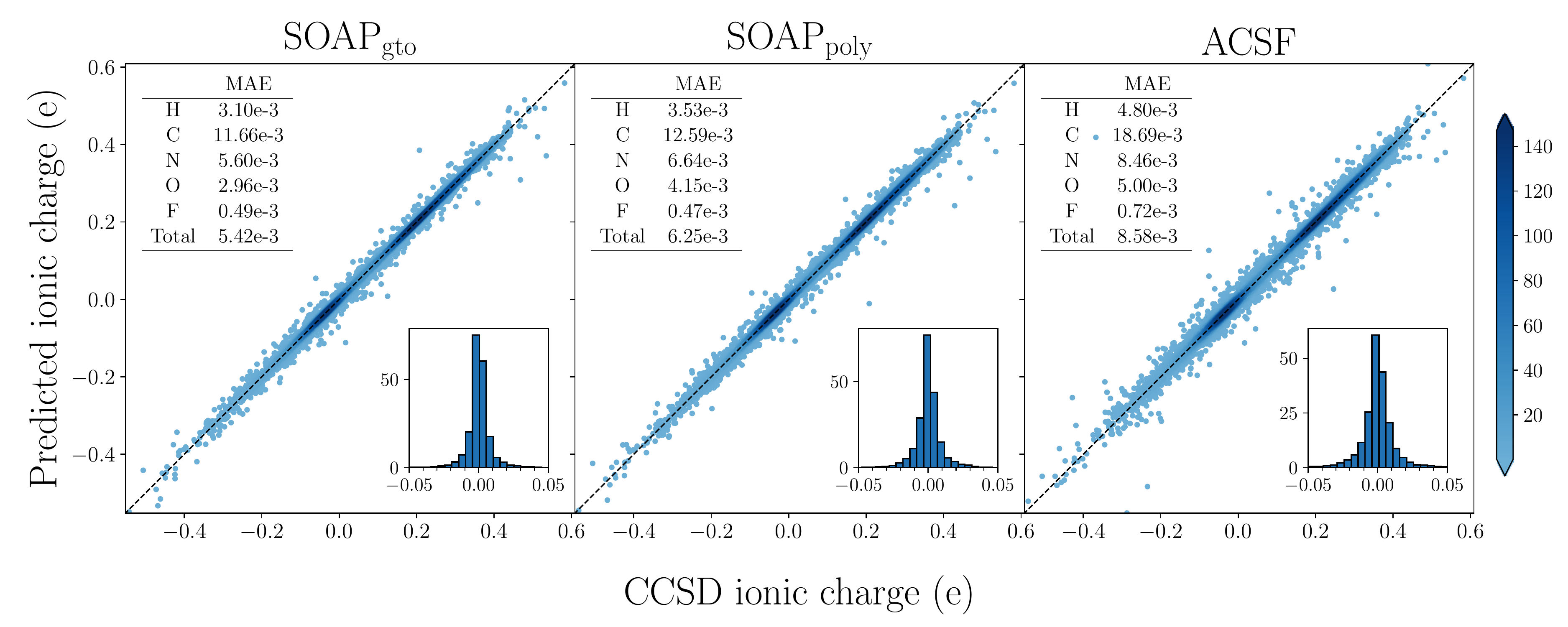}
  \caption{Parity plot of ionic charge prediction results from the test set against true CCSD ionic charges. The predictions are performed with kernel ridge regression using SOAP$_\text{gto}$ (gaussian type orbital basis), SOAP$_\text{poly}$ (polynomial basis) and ACSF. The mean absolute errors for the different elements is shown in the inset together with the total error distribution.}
  \label{fig:charge_prediction}
\end{figure*}

To demonstrate the prediction of local properties with the DScribe package, a prediction of ionic charges in small organic molecules is performed. The dataset consists of Mulliken charges calculated at the CCSD level for the GDB-9 dataset of 133 885 neutral molecules \cite{134k}. The structures contain up to nine atoms and five different chemical species: hydrogen, oxygen, carbon, nitrogen and fluorine. The geometries have been relaxed at the B3LYP/6-31G(2df,p) level and no significant forces were present in the static CCSD calculation.

The prediction is performed with the two local descriptors included in the package, SOAP and ACSF. For SOAP we perform the prediction with both radial basis functions: the polynomial basis (SOAP$_\text{poly}$) and the gaussian type orbital radial basis (SOAP$_\text{gto}$). For them we fix $n_\mathrm{max} = 8$ and $l_\mathrm{max} = 8$, but optimize the cutoff $r_\mathrm{cut}$ and Gaussian width $\sigma$ with grid search. For ACSF we use 10 radial functions $G^2$ and 8 angular functions $G^3$. The cutoff value $r_\mathrm{cut}$ is shared between the radial and angular functions and it is optimized with grid search. More details about the used ACSF symmetry functions are found in the Supplementary Information.

Descriptor hyperparameters are optimized with grid search separately for each species on a smaller set of 2500 sample atoms with 5-fold cross-vali\-da\-tion and 80\%/20\%-training/test split. Both the KRR kernel width and the regularization parameter are allowed to vary on a logarithmic grid during the descriptor hyperparameter search. After optimizing the descriptor hyperparameters, the model is retrained using the optimal descriptor setup and a larger dataset of 10 000 sample atoms for each species (except fluorine, for which only 3314 atoms were available in the dataset). The training is done with the same cross-validation setup as for the descriptor hyperparameter optimization, but now with finer grid for the KRR kernel width. The parity plots for the charge prediction are shown in Figure \ref{fig:charge_prediction}. The hyperparameter grids and optimal values for both the descriptors and kernel ridge regression are found in the Supplementary Information together with additional details.

\subsection{Discussion}
The formation energy prediction demonstrates that our implementation performs consistently and offers insight into the performance of the different descriptors. Special care must be taken in interpreting the results, as there exist different variations of the different descriptors. For example, as discussed in section \ref{sec:local_to_global}, there are different ways to combine information from multiple local SOAP-outputs, and different geometry functions and cutoff types may be used for the MBTR. The learning rates also depend on the chosen machine learning model.

Our results for the Ewald sum matrix and the sine matrix reflect the results reported earlier, where a formation energy prediction was performed for a similar set of data from the Materials Project \cite{materialsproject}. They report MAE for the Ewald sum matrix to be 0.49 eV and for the sine matrix to be 0.37 eV \cite{sm} with a training set of 3000 samples, whereas we find MAE for the Ewald sum matrix to be 0.36 eV and for the sine matrix to be 0.24 eV with a training set of 3276 samples. The performance improvement in our results can be explained by differences in the contents of the used dataset. We, however, recover the same trend of the sine matrix performing better, even when issues in the original formulation of the Ewald sum matrix (as discussed in section \ref{sec:ewald}) were addressed. The low performance of the more accurate charge interaction in the Ewald model and the relatively small difference between the performance of the Coulomb and sine matrix may indicate that for this task the information of the potential energy of the neutral atoms -- contained on the diagonal of both the sine and Coulomb matrix -- largely controls the performance.

With respect to the individual performance of the different MBTR parts, the $k=2$ terms containing distance information performs best, whereas the angle information contained in $k=3$ and the simple composition information contained by $k=1$ lag behind. However, the best MBTR performance is achieved by combining the information from all of the terms. It is also surprising how well the simple averaging scheme for SOAP performs in the tested dataset range. When extrapolating the performance to larger datasets, it can however be seen that MBTR may provide better results.

The charge prediction test illustrates that the ionic charges of different species in organic molecules may be learned accurately on the CCSD level just by observing the local arrangement of atoms up to a certain radial cutoff. On average the mean absolute error is around 0.005-0.01 $e$ when using a dataset of 10 000 samples for each species. The variance of the Mulliken charges in the training set depends on the species, which also results in species-specific variation of MAE for the predicted charges. As to be expected, the charge of the multi-valent species -- C, N, O -- varies much more in the CCSD data and is much harder to predict than the charge of the low valence species H and F.

Our comparison shows that there is little difference between the predictive performance of the two radial bases used for SOAP. With our current implementation there is, however, a notable difference in the speed of creating these descriptors. For identical settings ($n_\mathrm{max} = 8$, $l_\mathrm{max} = 8$, $r_\mathrm{cut} = 5$, and $\sigma$ = 0.1), the gaussian type orbital basis is over four times faster to calculate than the polynomial basis. This difference originates largely from the numerical radial integration, which is required for the polynomial basis but not for the gaussian type orbital basis. The prediction performance of ACSF does not fall far behind SOAP and it might be possible to achieve the same accuracy by using a more advanced parameter calibration for the symmetry functions. The symmetry functions used in ACSF are easier to tune for capturing specific structural properties, such as certain pairwise distances or angles formed by three atoms. This tuning can, however, be done only if such intuition is available \emph{a priori}, and in general consistently improving the performance by changing the used symmetry functions can be hard.

%===============================================================================
% Conclusions
%===============================================================================
\section{Conclusions}
The recent boom in creating machine learnable fingerprints for atomistic systems, or descriptors, has led to a plethora of available options for materials science. The software implementations for these descriptors is, however, often scattered across different libraries or missing altogether, making it difficult to test and compare different alternatives.

We have collected several descriptors in the DScribe software library. DScribe has an easy-to-use python-interface, with C/C++ extensions for the computationally intensive tasks. We use a set of regression tests to ensure the validity of the implementation, and provide the source code together with tutorials and documentation. We have demonstrated the applicability of the package with the supervised learning tasks of formation energy prediction for crystals and the charge prediction for molecules. The DScribe descriptors are compatible with general-purpose machine learning algorithms, and can also be used for unsupervised learning tasks. In the future we plan to extend the package with new descriptors, such as the structural descriptor based on Voronoi tesselations \cite{voronoi}, and also welcome external contributors.

%================================================================================
% Acknowledgements
%================================================================================
\section{Acknowledgements}
We acknowledge the computational resources provided by the Aalto Science-IT project. This project has received funding from the Jenny and Antti Wihuri Foundation and the European Union's Horizon 2020 research and innovation programme under grant agreements number no. 676580 NOMAD and no. 686053 CRITCAT.

%===============================================================================
% Appendix


%===============================================================================
% References


\bibliographystyle{elsarticle-num-names}
\biboptions{numbers,sort&compress}
\bibliography{bibliography}



%\begin{thebibliography}{00}
%  \input{bibliography.bbl}
%\end{thebibliography}

\end{document}